\documentclass[12pt]{iopart}
\usepackage{graphicx}
\begin{document}

\title{Method to reduce excess noise of a detuned cavity for application in KAGRA}
\author{Shinichiro Ueda$^1$, Nana Saito$^2$, Daniel Friedrich$^3$, Yoichi Aso$^4$, and Kentaro Somiya$^1$ (for the KAGRA collaboration)}
\address{$^1$Graduate School of Science and Technology, Tokyo Institute of Technology, 2-12-1 Oh-okayama, Meguro-ku, Tokyo, Japan}
\address{$^2$Graduate School of Humanities and Sciences, Ochanomizu University, 2-1-1 Otsuka, Bunkyo-ku, Tokyo, Japan}
\address{$^3$National Astronomical Observatory, 2-21-1 Osawa, Mitaka-shi, Tokyo, Japan}
\address{$^4$Department of Physics, University of Tokyo, 7-3-1 Hongo, Bunkyo-ku, Tokyo, Japan}
\ead{ueda.s@gw.phys.titech.ac.jp}

\begin{abstract}

Ground-based gravitational-wave detectors are based on high precision laser interferometry.
One promising technique to improve the detector's sensitivity is the detuning of an optical cavity, which enhances the signal at around certain frequencies for target astronomical sources. The detuning, however, involves technical noise due to an asymmetry of the control sidebands, which includes photo-detector noise and oscillator-phase noise. Here, we introduce a solution to reduce the two kinds of excess noise using an amplitude-modulation sideband that compensates the asymmetry. The solution is planned to be implemented in the Japanese second-generation gravitational-wave detector KAGRA.
\end{abstract}


\maketitle

\section{Introduction} \label{sec:1}
Currently, large scale interferometric gravitational-wave detectors, Advanced LIGO~\cite{LIGO}, Advanced Virgo~\cite{Virgo}, GEO-HF~\cite{GEO} and KAGRA~\cite{KAGRA}, are in the middle of an upgrade or under construction. Those second-generation detectors introduce new technologies to improve their sensitivities. Among these, a promising technique to decrease quantum noise is the detuning of the signal-recycling cavity.

\begin{figure}[htbp]
\begin{center}
  \includegraphics[width=9cm]{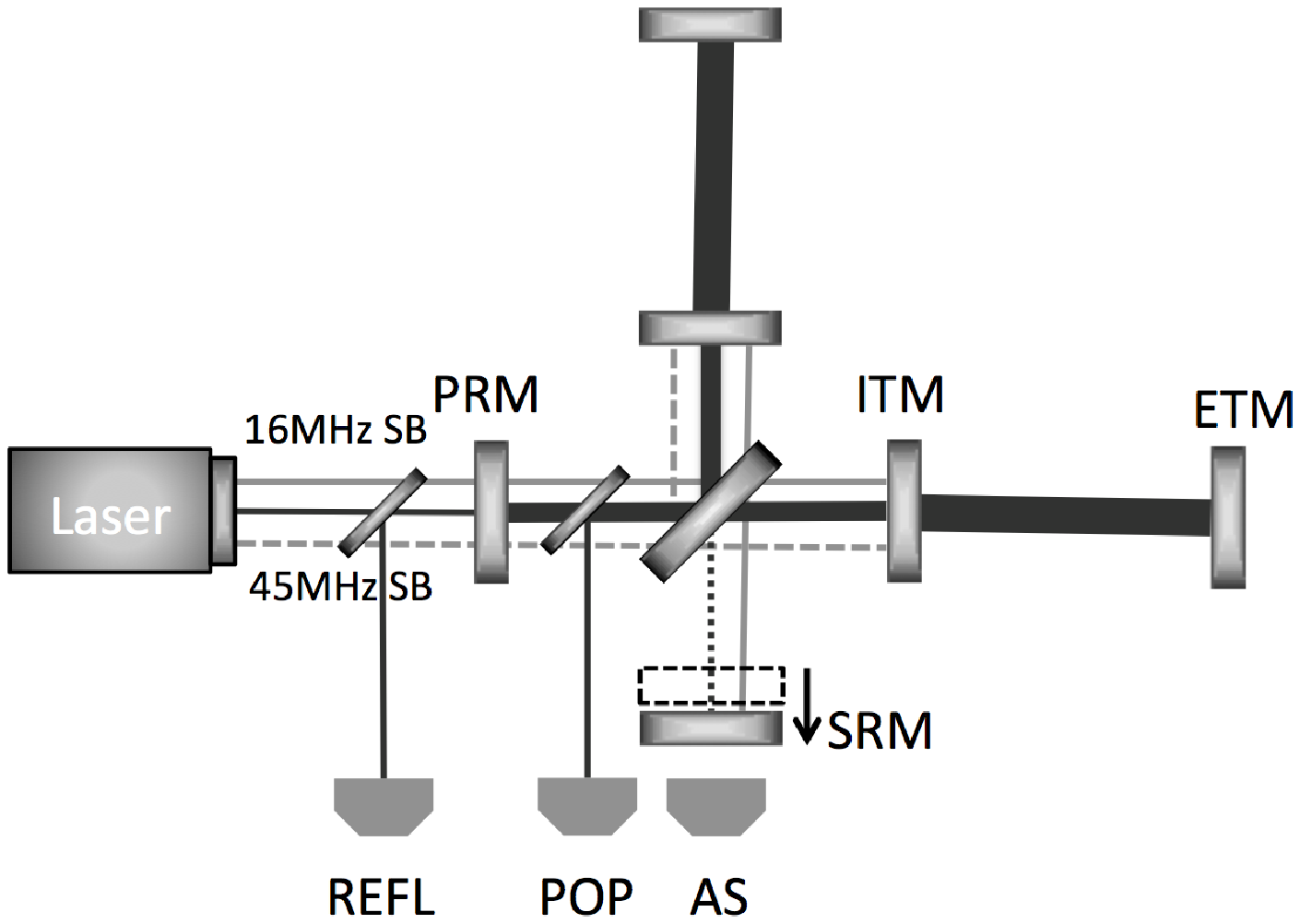}
 \caption{Simplified schematic of KAGRA in the detuned configuration. The black solid, gray solid and gray dashed lines express the carrier, a 16\,MHz sidebands, and a 45\,MHz sidebands, respectively. Since there are five degrees of freedom to be controlled, the two control sidebands and three signal extraction ports are necessary. The gravitational-wave signal is extracted from the antisymmetric port (AS) using the carrier light. The other control signals are extracted by taking a beat of the carrier and one of the sidebands~\cite{Aso}.}
 \label{Conf_DRSE}
\end{center}
\end{figure}

The optical configuration of KAGRA is the so-called resonant sideband extraction (RSE) topology, which is based on a Michelson interferometer with (i) a Fabry-Perot cavity in each arm, (ii) a resonant power-recycling cavity, and (iii) an anti-resonant or slightly detuned signal-recycling cavity~\cite{Mizuno} as sketched in Fig.~\ref{Conf_DRSE}. In the 'tuned' RSE, the signal-recycling cavity is locked to the anti-resonance of the carrier light so that the signal is increased at higher frequencies at the expense of a decreased sensitivity at low frequencies. In the detuned RSE, the signal-recycling cavity is slightly detuned from the anti-resonance so that the signal at around a certain frequency band is increased~\cite{BC}. This technique only improves the quantum noise level in a narrow frequency band while it degrades it at other frequencies. However, the observation rate can be increased for binary neutron star merger events, which are the primary target of KAGRA.
Recently, we have found that the detuning can increase the couplings of other noises, which includes photo-detector noise (PDN) and oscillator phase noise (OPN).  These noise sources could make the sensitivity of detuned RSE actually worse when compared to the case of tuned RSE.

In this paper, we propose a method to address the increased coupling of the PDN and OPN using an additional amplitude modulation (AM) sidebands. In Sec.~\ref{sec:2}, we explain the length sensing scheme of KAGRA. In Sec.~\ref{sec:3}, we show the mechanism of the noise increase with the detuning. In Sec.~\ref{sec:4}, we introduce our method to reduce the couplings of the PDN and OPN. In Sec.~\ref{sec:5} we show our experimental demonstration of the optical detuning with a single Fabry-Perot cavity. In Sec.~\ref{sec:6}, we present results of our simulation to confirm the noise reduction and the estimated performance based on the parameters of KAGRA. Finally in Sec.~\ref{sec:7}, we summarize the study. 

\section{Length sensing scheme of KAGRA} \label{sec:2}

\begin{figure}[htbp]
\begin{center}
  \includegraphics[width=9cm]{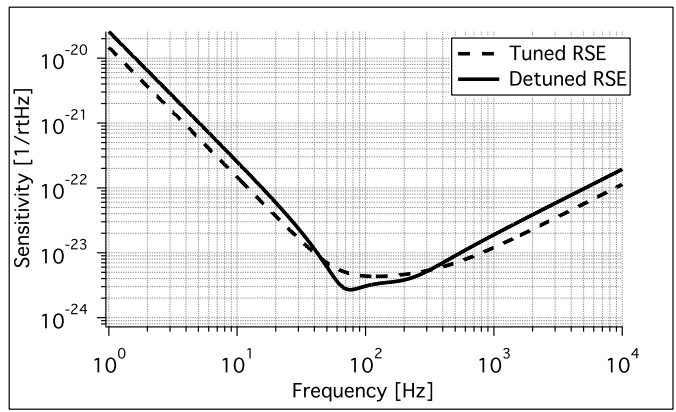}
 \caption{Quantum noise spectra of KAGRA in the tuned (dashed) and detuned (solid) RSE configurations. In the detuned RSE, the signal-recycling mirror position is shifted from the anti-resonance by $\sim$10~nm. The sensitivity of the detuned RSE is improved at around 100~Hz compared with the tuned RSE.}
 \label{sense_RSEvsDRSE}
\end{center}
\end{figure}
Figure~\ref{Conf_DRSE} shows the detuned RSE configuration. There are three signal extraction ports named antisymmetric (AS), reflection (REFL), and pick-off in the power-recycling (POP).
In the case of the tuned RSE, the signal-recycling cavity does not further recycle but extracts the gravitational-wave signal at high frequencies before it starts to over-circulate in the arm cavities. As a result, by improving the sensitivity at higher frequencies the observation band is broadened.
 
In each arm of the tuned RSE, the input and end mirrors stay in the position where the radiation pressure of the laser and restoring forces of the pendulums that suspend the test masses are balanced (control forces are also applied to the test masses in reality).
Changing the control force, we can shift the balanced position. A radiation pressure proportional to the displacement from the new balanced position is applied to the test masses and it creates some sort of a spring, which we call {\it optical spring}~\cite{BC}.
In the gravitational-wave detector, the central Michelson interferometer is locked to the dark fringe to separate signals for the common and differential motions of the test masses.
The optical spring of the differential signal is not produced by shifting the test-mass positions in the arm cavities but by shifting the the signal-recycling mirror position.
This interferometer configuration is called detuned RSE. 
Figure~\ref{sense_RSEvsDRSE} shows sensitivity spectra of the tuned and detuned RSE with KAGRA parameters. The detuned RSE shows a better sensitivity at around 100\,Hz.

KAGRA employs two phase modulation (PM) sidebands, one at 16\,MHz (gray solid line in Fig.~\ref{Conf_DRSE}) and the other at 45\,MHz (gray dashed line) besides the carrier light (black solid line). The 16\,MHz sidebands are circulating in both the power-recycling and signal-recycling cavities while the 45\,MHz sidebands circulate only in the power-recycling cavity. The 16\,MHz sidebands are used to extract the information of the signal-recycling mirror motion for its position control.

\begin{figure}[htbp]
\begin{center}
  \includegraphics[width=7cm]{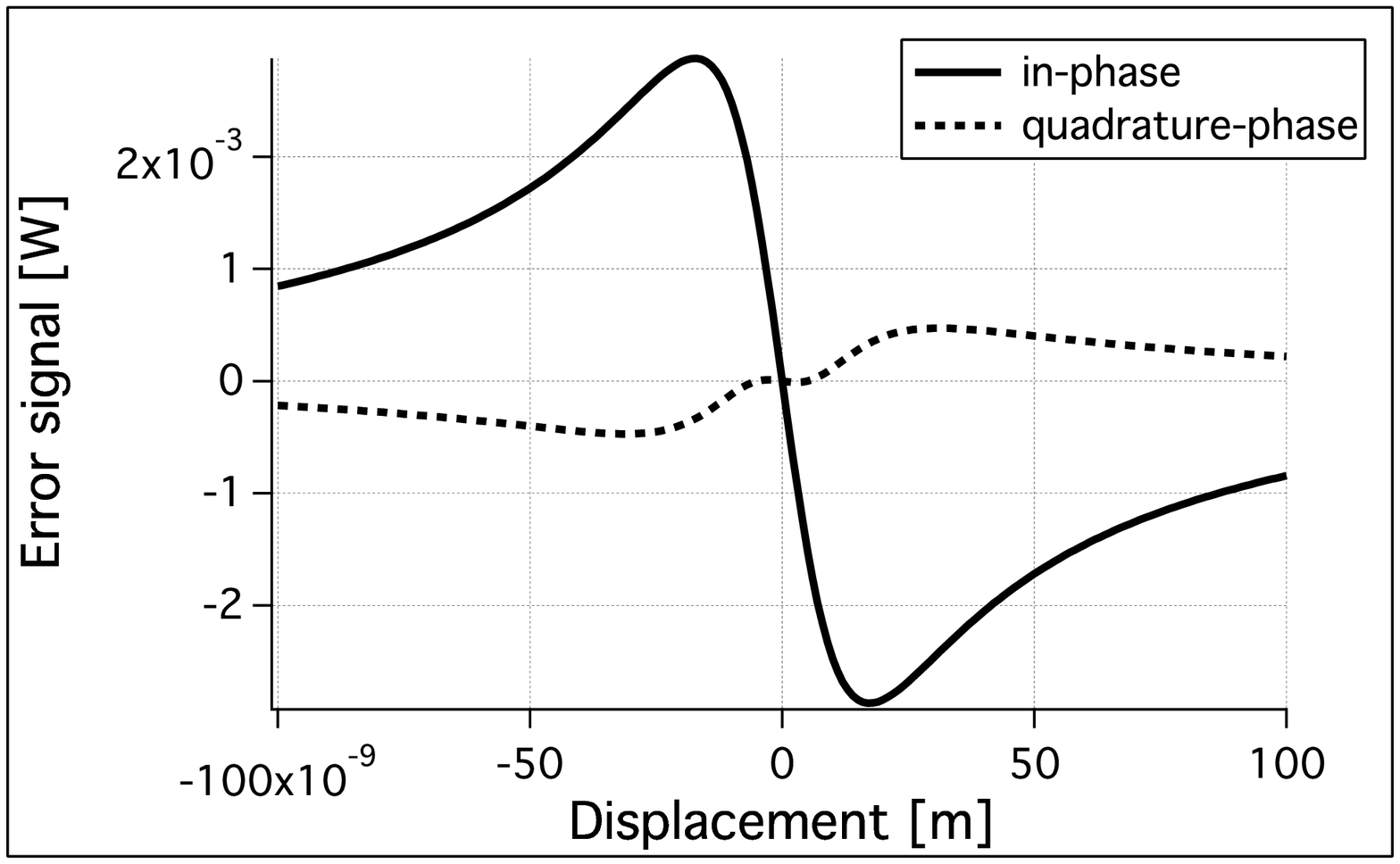}
  \includegraphics[width=7cm]{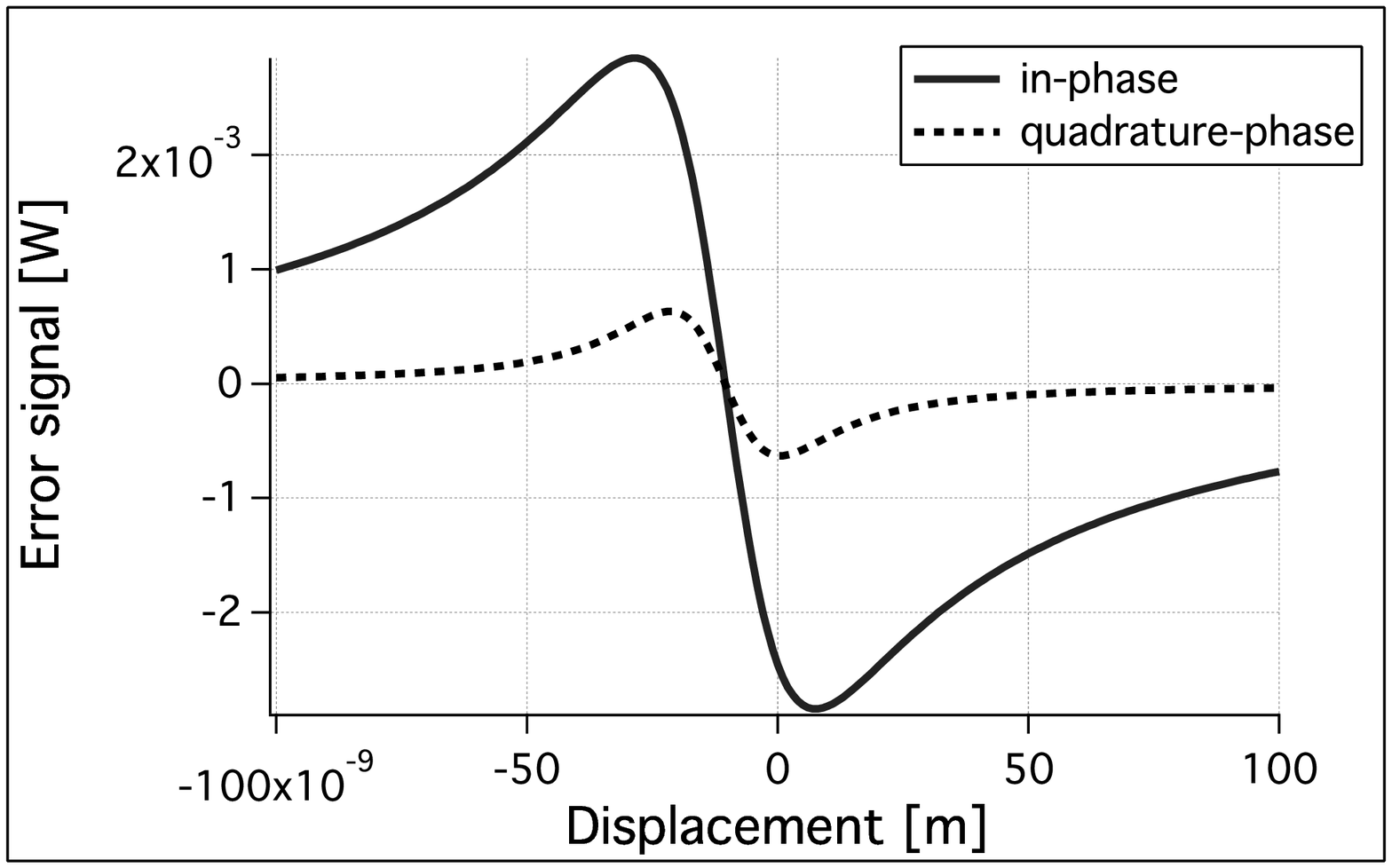}
\end{center}
 \caption{{\it Left}: Error signal of SRCL without detuning. {\it Right}: Error signal with detuning. The error signal has an offset of $-2.5\times 10^{-3}$~W at the operating point.}
 \label{error-signal}
\end{figure}

In order to lock each mirror to its operating point, we should obtain a proper error signal, which tells us how much the mirror is displaced. The left panel of Fig.~\ref{error-signal} shows an error signal of the signal-recycling mirror in KAGRA. The error signal shows a linear response to the mirror displacement at around the zero point, which is the operation point of the mirror. By feeding back the error signal to the mirror with a coil-magnet or electrostatic actuator, the mirror can stay at an appropriate position.

The RSE system has five longitudinal degrees of freedom to be controlled and thus we need five error signals. These five degrees of freedom are called differential arm length (DARM), common arm length (CARM), central Michelson interferometer (MICH), power recycling cavity length (PRCL) and signal-recycling cavity length (SRCL). The DARM (CARM) signal expresses the differential (common) motion of the two arm cavities, MICH is the differential motion of the lengths between the beamsplitter and the two input test masses, and PRCL/SRCL are the motions of the power/signal-recycling mirrors. Information of DARM is extracted from the AS with the carrier and other information is extracted from POP or REFL with the combination of the carrier and the 16\,MHz or 45\,MHz sidebands. The 16\,MHz sidebands are used for SRCL as it is the only electric field that contains some information of the signal-recycling mirror motion.

\section{Excess noise caused by the detuning} \label{sec:3}

In this section, we discuss the problems associated with the detuning of the signal-recycling cavity. 
The detuning causes some kinds of excess noise. We mainly discuss photo-detector noise (PDN) and oscillator phase noise (OPN), which are the most serious problems among them.
First, we shall explain a phasor diagram, which is used in the following discussions.
The phasor diagram expresses electric fields in a complex plane with the carrier light $E_0e^{\mathrm{i}\Omega t}$ fixed on the horizontal axis, or AM-axis. A phase-modulation sidebands to the carrier light appear on the vertical axis, or the PM-axis. In the left panel of Fig.~\ref{phasor_dia}, the electric fields are represented by arrows. The amplitude of the field and the relative phase to the carrier is represented by the length and the argument of the arrow, respectively, in the complex plane.
The output field after the phase modulation is described as 
\begin{eqnarray}
E(t)&=&E_0J_0(\mathrm{m})e^{\mathrm{i}\Omega t} + \mathrm{i}E_0J_1(\mathrm{m})\left[e^{\mathrm{i}(\Omega+\omega) t} + e^{\mathrm{i}(\Omega-\omega) t}\right]\ ,\label{eq:modulation}
\end{eqnarray}
where $m$ is modulation depth, $\Omega$ is the frequency of the carrier, $\omega$ is the frequency of the PM-sidebands, and $J_\mathrm{n}$ is the n-th order Bessel function.
The first term of Eq.~(\ref{eq:modulation}) is the carrier light, and the second and the third terms are the upper and lower sidebands that rotate in opposite directions in the phasor diagram (Fig.~\ref{phasor_dia}, left panel). Combining the upper and lower sidebands, one can see that the phase-modulation sidebands oscillate up and down on the vertical axis. The signal obtained with both the upper and lower sidebands pointing toward the same direction is called an {\it in-phase} signal and the signal with both pointing toward the opposite directions is called a {\it quadrature-phase} signal.

\subsection{Photo-detector noise} \label{sec:PDN}
\begin{figure}[htbp]
  \begin{center}
  \includegraphics[width=7cm]{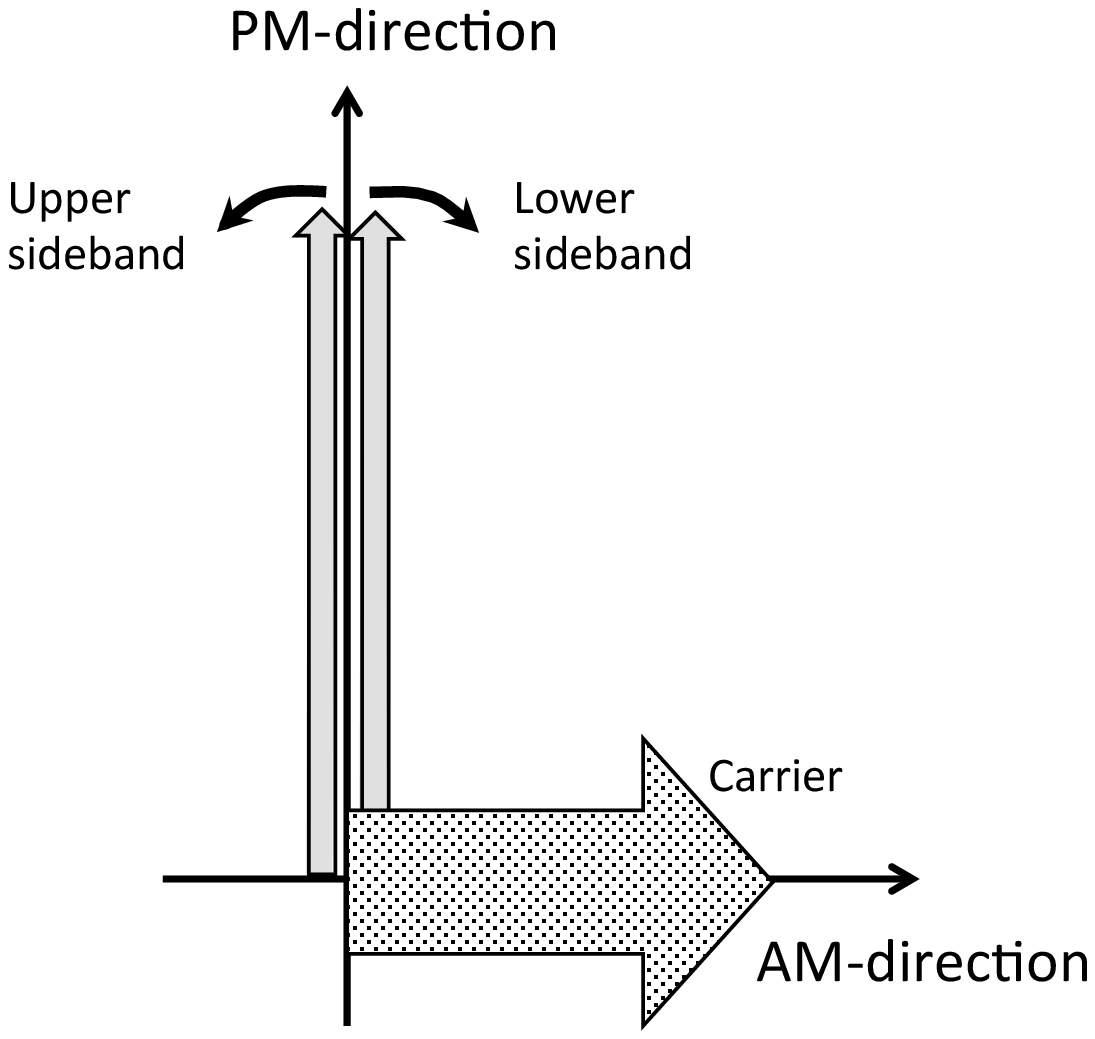}
  \includegraphics[width=6cm]{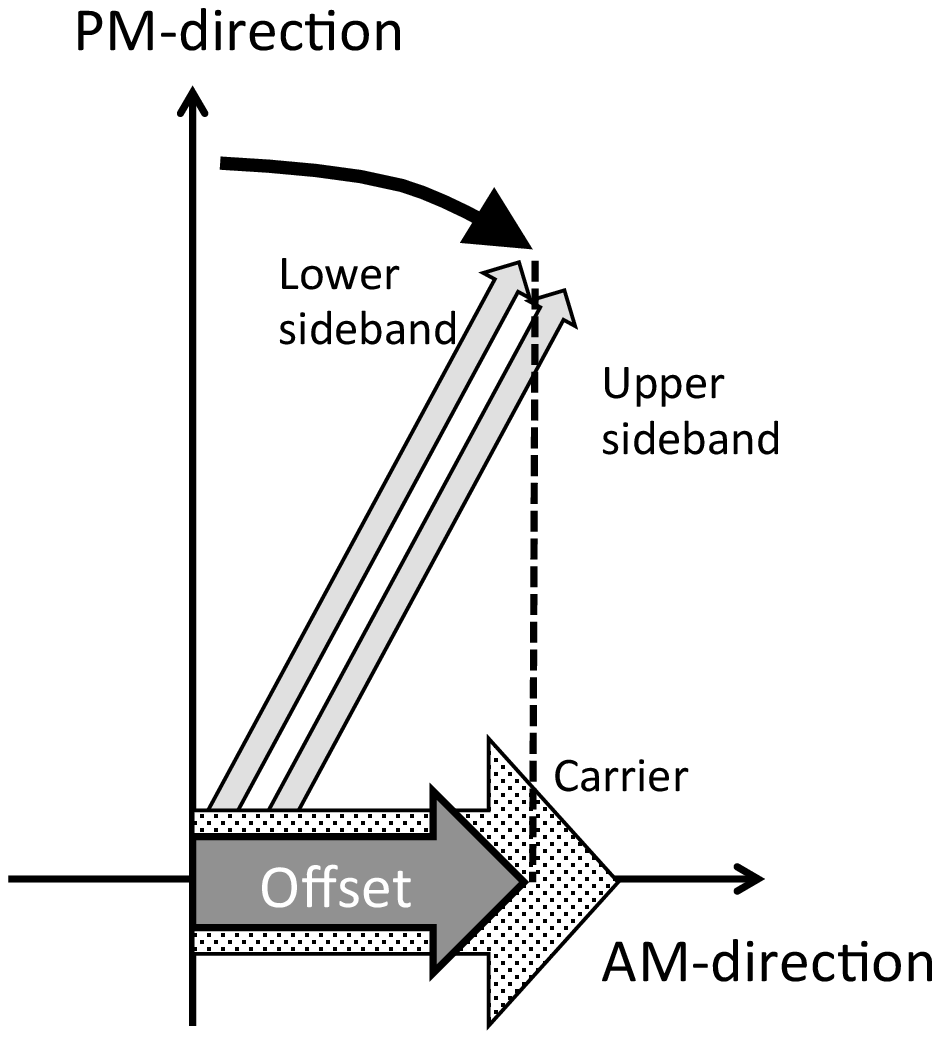}
 \end{center}
 \caption{{\it Left}: A phasor diagram for a modulated electric field. The carrier (colored gray) is directed to the right and the upper and lower sidebands (colored light gray) rotate in opposite directions at the modulation frequency. For the phase modulation, the sum of the upper and lower sidebands oscillate up and down on the vertical axis. For the amplitude modulation, it oscillates right and left on the horizontal axis. {\it Right}: The upper and lower sidebands of the PM sidebands (light gray arrows) rotate together by a same amount in the phasor diagram due to the detuning. Its non-zero projection onto the AM-axis (gray arrow) beats with the carrier and makes a 16\,MHz flashing, which turns into an offset on the error signal after the demodulation.}
 \label{phasor_dia}
\end{figure}

A photo detector has a finite dynamic range. With a large voltage introduced on the photo detector, the gain of the detector that gives the conversion efficiency of the signal must be lowered to avoid the saturation of the readout. In other words, the offset voltage decreases the resolution of the photo detector. This happens at around each RF sideband frequency that is used to extract a control signal of the interferometer.
A monotone offset voltage at a certain frequency on the photo-detector decreases the resolution and thus increases PDN.
In the case of the detuned RSE, the 16\,MHz offset light appears at REFL and POP ports.
The right panel of Fig.~\ref{error-signal} shows how much offset is generated after the demodulation at 16~MHz. Both the upper and lower 16~MHz sidebands rotate a little from the PM-axis to the AM-axis due to the detuning of the signal-recycling cavity. Coupling with the carrier, a fraction of the sidebands projected to the AM-axis makes a 16~MHz flashing, which turns to an offset voltage after the demodulation. A typical value of resolution noise due to the finite dynamic range is about 1\,nV for a 100\,mV output, and then the PDN level with an offset is given as follows~\cite{LSC}:
\begin{eqnarray}
\mathrm{PDN~(m/\sqrt{\mathrm{Hz}})} = \frac{\mathrm{offset (W)}}{\mathrm{optical\ gain (W/m)}} \times \frac{1\mathrm{nV/\sqrt{\mathrm{Hz}}}}{100\mathrm{mV}}\ . \label{eq:PDN}
\end{eqnarray}
The right panel of Fig.~\ref{phasor_dia} shows the offset in the SRCL error signal (in-phase). With the detuning, a beat of the offset and carrier generates the 16\,MHz flashing on the photo detector. Demodulated at 16\,MHz, the flashing makes a DC offset that appears on the error signal. 
In a simulation with the parameters for KAGRA (see Table~\ref{tb:prKAGRA}), the amount of the offset is $-2.5\times 10^{-3}$~W. 
In the noise spectrum, PDN with the offset is as large as 1/3 of the KAGRA target sensitivity (Fig.~\ref{PDNandOPN}). 

\subsection{Oscillator phase noise} \label{sec:OPN}

\begin{figure}[htbp]
\begin{center}
\includegraphics[width=8cm]{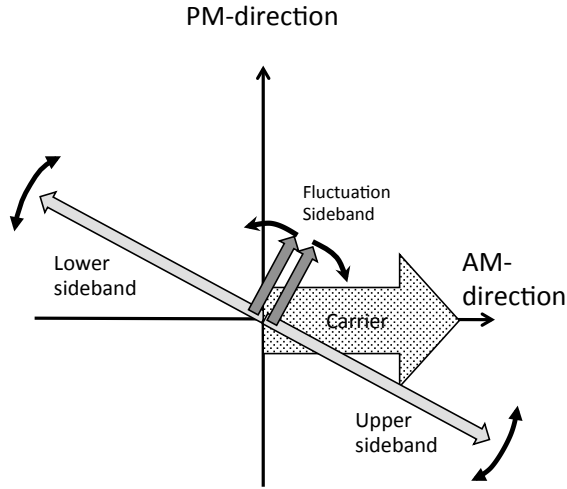}
\end{center}
\caption{Oscillator phase noise fluctuation sidebands (gray arrows) in the quadrature-phase signal. The arrow with dotted hatching is the carrier and the light gray arrows are the upper and lower 16\,MHz sidebands.}
\label{OPN}
\end{figure}

Oscillators that generate a modulation to the carrier light via electro-optic modulators (EOMs) imposes noise on sidebands. There are two types of noise, namely amplitude noise and phase noise. Typically the latter, namely OPN, can be more problematic in an interferometer~\cite{Aso}.
OPN is represented by fluctuation sidebands on the upper and lower 16\,MHz sidebands. For the in-phase signal, the fluctuation sidebands are facing to opposite directions and cancel out as far as the upper and lower 16\,MHz sidebands are balanced. For the quadrature-phase signal, the fluctuation sidebands are facing to the same direction and add up.
As we have explained in Sec.~\ref{sec:PDN}, the 16\,MHz sidebands are tilted in the complex plane due to the detuning. Therefore the fluctuation sidebands can couple with the carrier to produce noise directly in the gravitational-wave channel (Fig.~\ref{OPN}).

\subsection{Imbalance between upper and lower sidebands} \label{sec:Imbarance}
\begin{figure}[htbp]
 \begin{center}
  \includegraphics[width=15cm]{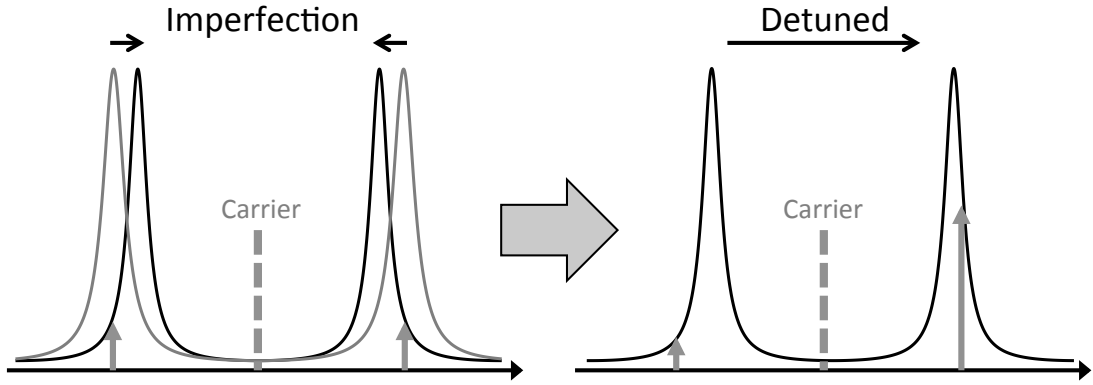}
 \end{center}
 \caption{In tuned RSE, the imperfection of the beamsplitter changes the free spectral range of the signal-recycling cavity (from the gray curve to the black curve) and the recycling gain is lowered by the same amount for both the upper and lower sidebands (gray arrows in the left panel). In detuned RSE, the black curve shifts to one direction and the recycling gain becomes different between the sidebands to make an imbalance (gray arrows in the right panel).}
 \label{imbalanceofBS}
\end{figure}

So far we have assumed that the upper and lower sidebands are balanced. However, an imperfection of the beamsplitter, namely a mismatch in the reflectivity and transmittance, can cause an imbalance between the amplitude of the upper and lower sidebands. 
Let us first consider the tuned RSE, shown in the left panel of Fig.~\ref{imbalanceofBS}. Without the imperfection of the beamsplitter, both the $\pm$16\,MHz sidebands are on resonance (gray curve). With the imperfection, the effective signal-recycling cavity length changes and the upper and lower sidebands are slightly off resonant by the same amount. Now let us think of the detuned RSE, shown in the right panel of Fig.~\ref{imbalanceofBS}. With the slight off resonance due to the imperfection, either the upper or lower sideband comes closer to the resonance with the detuning. The difference of the recycling gain results in an imbalance of the amplitudes between the upper and lower sidebands.

\section{Solution with AM sidebands} \label{sec:4}

The two types of excess noise discussed above originate from the tilted pair of upper and lower 16\,MHz sidebands due to the detuning. Both types of noise can be reduced when the tilt from the PM-axis is compensated. Such a situation can be realized by adding 16\,MHz amplitude modulation sidebands beforehand. Here, we assume an electro-optic modulator for the AM sidebands added in series to the modulator for the PM sidebands.

In the tuned RSE, the electric field reflected from the interferometer can be written as
\begin{eqnarray}
E^\mathrm{T} = E_0^\mathrm{T} \mathrm{e}^{\mathrm{i}\Omega t} + \mathrm{i}E_1^\mathrm{T}\mathrm{e}^{\mathrm{i}(\Omega+\omega) t} +\mathrm{i} E_1^\mathrm{T}\mathrm{e}^{\mathrm{i}(\Omega-\omega) t}\ ,  \label{woAM}
\end{eqnarray}
where the subscripts $0$ and $1$ indicate the amplitudes of the carrier and the PM sidebands reflected from the interferometer, respectively. In the detuned RSE, the electric field reflected from the interferometer is then given by
\begin{eqnarray}
E^\mathrm{D} = E_0^\mathrm{D} \mathrm{e}^{\mathrm{i}\Omega t} + \left(\mathrm{i}E_{+1}^\mathrm{D}\mathrm{e}^{\mathrm{i}(\Omega+\omega) t} +\mathrm{i} E_{-1}^\mathrm{D}\mathrm{e}^{\mathrm{i}(\Omega-\omega) t}\right)\mathrm{e}^{\mathrm{i} \alpha},  \label{woAM}
\end{eqnarray}
where the subscripts $\pm1$ indicate the upper and lower sidebands of the PM sidebands, and $\alpha$ is the angle from the PM-axis in the phasor diagram when the upper and lower sidebands are pointing to the same direction.

\begin{figure}[htbp]
 \begin{center}
  \includegraphics[width=15cm]{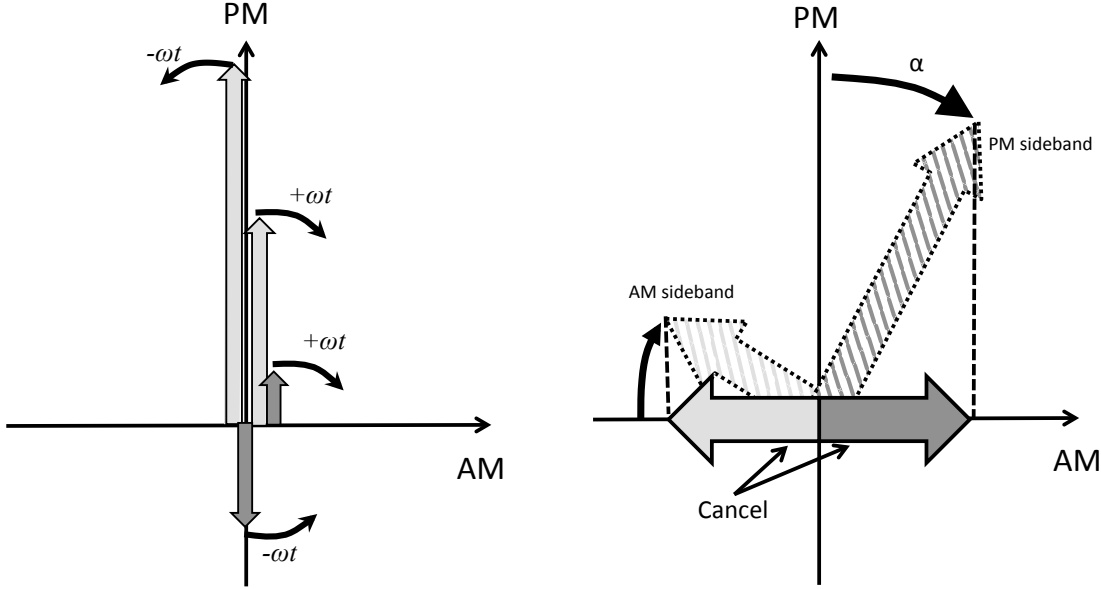}
 \end{center}
 \caption{{\it Left}: Compensation of the imbalance between the upper and lower sidebands of the PM sidebands using the additional AM sidebands. {\it Right}: Compensation of the tilt of the PM sidebands due to the detuning using the additional AM sidebands.}
 \label{cancel_imbalance}
\end{figure}

First, let us focus on the imbalance between the amplitudes of the upper and lower sidebands, assuming $\alpha = 0$. Adding AM sidebands at the relative phase of 90\,deg to the PM sidebands, one can balance the amplitudes of the upper and lower sidebands. The four electric fields can be divided into two pairs of sidebands at the modulation frequencies $\pm\omega$. At the relative phase of 90\,deg, one of the AM sidebands components increases and the other decreases the amplitude of the PM sidebands components as illustrated in the left panel of Fig.~\ref{cancel_imbalance}. Defining the amplitudes of the upper and lower sidebands of the AM sidebands as $E_{\pm1}^{\mathrm{AM}}$, one can describe the total field as
\begin{eqnarray}
E &= E_0^\mathrm{D} \mathrm{e}^{\mathrm{i}\Omega t}  + \mathrm{i}E_{+1}^\mathrm{D}\mathrm{e}^{\mathrm{i}(\Omega +\omega) t} +\mathrm{i} E_{-1}^\mathrm{D}\mathrm{e}^{\mathrm{i}(\Omega-\omega) t} \nonumber\\
&\hspace{45mm}+ E_{+1}^\mathrm{AM}\mathrm{e}^{\mathrm{i}((\Omega+\omega) t + \pi/2)} + E_{-1}^\mathrm{AM}\mathrm{e}^{\mathrm{i}((\Omega-\omega) t - \pi/2)} \nonumber \\
&= E_0^\mathrm{D} \mathrm{e}^{\mathrm{i}\Omega t}  + \mathrm{i}(E_{+1}^\mathrm{D} + E_{+1}^\mathrm{AM})\mathrm{e}^{\mathrm{i}(\Omega+\omega) t} +\mathrm{i} (E_{-1}^\mathrm{D} -E_{-1}^\mathrm{AM})\mathrm{e}^{\mathrm{i}(\Omega-\omega) t}.\label{wAMforUnbal}
\end{eqnarray}
Since the PM and AM sidebands will be equally influenced by the imperfection of the beamsplitter, the imbalance between the upper and lower sidebands of the PM and AM sidebands must satisfy
\begin{eqnarray}
\frac{E_{+1}^\mathrm{AM}}{E_{-1}^\mathrm{AM}} = \frac{E_{+1}^\mathrm{D}}{E_{-1}^\mathrm{D}}\ \ \ \left(\equiv a\right) . \label{eq:rateSB}
\end{eqnarray}
Choosing $E_{\pm1}^\mathrm{AM} = (a-1)/(a+1)E_{\pm1}^\mathrm{D}$, we can balance the amplitudes of the upper and lower sidebands.

Next, let us focus on the tilt of the PM sidebands in the phasor diagram, assuming the amplitudes of the upper and lower sidebands are equal: $E_{+1}^\mathrm{D} = E_{-1}^\mathrm{D} \equiv E_{\mathrm{1}}^{\mathrm{D}}$. Adding AM sidebands at the relative phase shift of 0\,deg to the PM sidebands, one can adjust the tilt without changing the amplitude balance. Reflected back from the detuned interferometer, both the PM and AM sidebands are tilted in the phasor diagram. With tuning the AM-sidebands amplitude $E_1^\mathrm{AM}$, one can cancel the AM-axis components of the PM and AM sidebands (Fig.~\ref{cancel_imbalance}, right panel).
The total field is described as
\begin{eqnarray}
E &= E_0^{\mathrm{D}} \mathrm{e}^{\mathrm{i}\Omega t}  + (\mathrm{i}E_1^{\mathrm{D}} + E_1^\mathrm{AM})\mathrm{e}^{\mathrm{i}\alpha}\mathrm{e}^{\mathrm{i}(\Omega+\omega) t} +(\mathrm{i}E_1^{\mathrm{D}} + E_1^\mathrm{AM})\mathrm{e}^{\mathrm{i}\alpha}\mathrm{e}^{\mathrm{i}(\Omega-\omega) t} \nonumber \\
&= E_0^{\mathrm{D}} \mathrm{e}^{\mathrm{i}\Omega t} + E_\mathrm{total}\mathrm{e}^{\mathrm{i}(\alpha + \beta)}\mathrm{e}^{\mathrm{i}(\Omega+\omega) t}+E_\mathrm{total}\mathrm{e}^{\mathrm{i}(\alpha + \beta)}\mathrm{e}^{\mathrm{i}(\Omega-\omega) t}\ ,\label{wAMforNoise}
\end{eqnarray}
where we have defined
\begin{eqnarray}
\beta = \arctan{\left[E_1^{\mathrm{D}}/E_1^\mathrm{AM}\right]},\ \ \ 
E_\mathrm{total} = \sqrt{(E_1^\mathrm{D})^2 + (E_1^\mathrm{AM})^2}\ .
\end{eqnarray}
If $E_1^\mathrm{AM}$ is chosen to satisfy $\alpha + \beta = \pi /2$, the total field becomes a pure PM sidebands:
\begin{eqnarray}
E= E_0 \mathrm{e}^{\mathrm{i}\Omega t} + \mathrm{i}E_{\mathrm{total}}\mathrm{e}^{\mathrm{i}(\Omega+\omega) t} +\mathrm{i}E_{\mathrm{total}}\mathrm{e}^{\mathrm{i}(\Omega-\omega) t}. \label{wAMfinal}
\end{eqnarray}

In reality, the PM sidebands are both unbalanced and tilted in the phasor diagram, but it is still possible to recover a pure PM balanced sideband by adding AM sidebands in a proper way. More details are given in \ref{sec:cal}.

\section{Experimental demonstration} \label{sec:5}

Let us demonstrate our method to create an optical offset using the mixed modulation in a single cavity experiment. While our goal for the gravitational-wave detector is to cancel the offset due to the detuning of the signal-recycling cavity on other degrees of freedom, the fundamental concept is to optically create the offset using AM sidebands. The purpose of our experiment here is to demonstrate the creation of the offset on the control signal in a single cavity by mixing the amplitude modulation in the phase modulation.

\begin{figure}[htbp]
	\begin{center}
		\includegraphics[width=10cm]{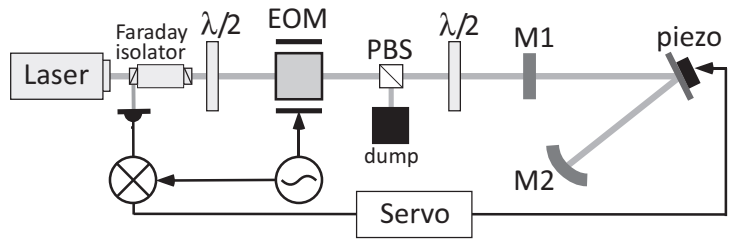}
	\caption{Experimental setup. Rotating the polarization of the beam at the first half-wave plate, one can create amplitude modulation together with the phase modulation. The amplitude modulation component imposes an offset in the error signal of the folded Fabry-Perot cavity.}
	\label{fig:setup}
	\end{center}
\end{figure}

Figure~\ref{fig:setup} shows a sketch of the experimental setup. We use a 1064\,nm continuous-wave laser by LIGHTWAVE with an output power of 40\,mW. There are two half-wave plates in between the laser and the cavity. The first half-wave plate is used to generate the amplitude modulation with a modulator and a polarization beam splitter (PBS). The second half-wave plate is used to recondition the shape of the laser beam, which is deformed after the PBS. This deformation can be caused by birefringence. The folded Fabry-Perot cavity consists of three mirrors. M1 is a flat mirror with a reflectivity of 98~\%, M2 is a curved high-reflectivity mirror, and a steering mirror is attached to a piezo actuator. The modulator is a resonant-type EOM at 12\,MHz. The reflected light is separated by means of a Faraday isolator.

\begin{center}
\begin{figure}[t]
  \includegraphics[width=7cm]{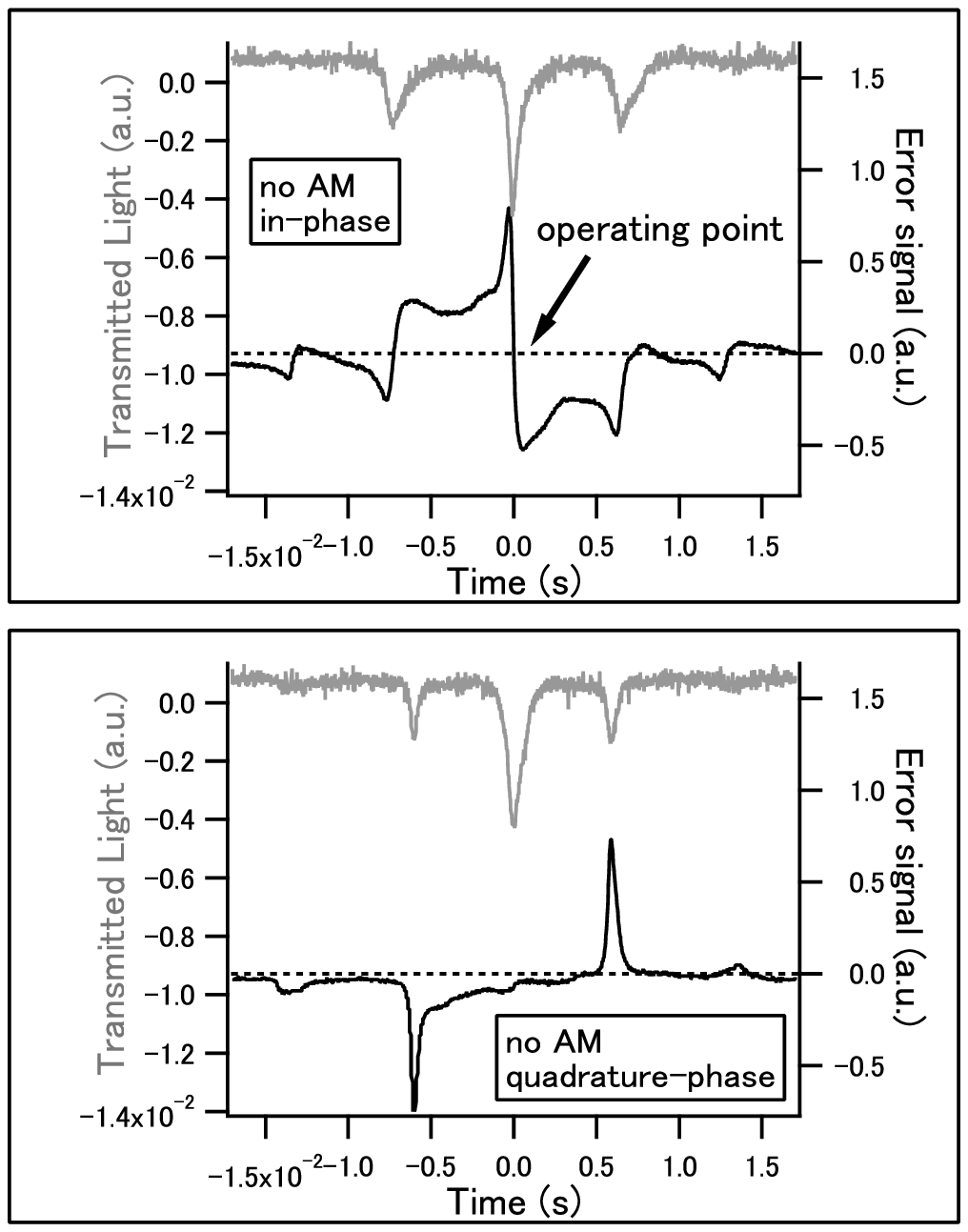}
  \includegraphics[width=7cm]{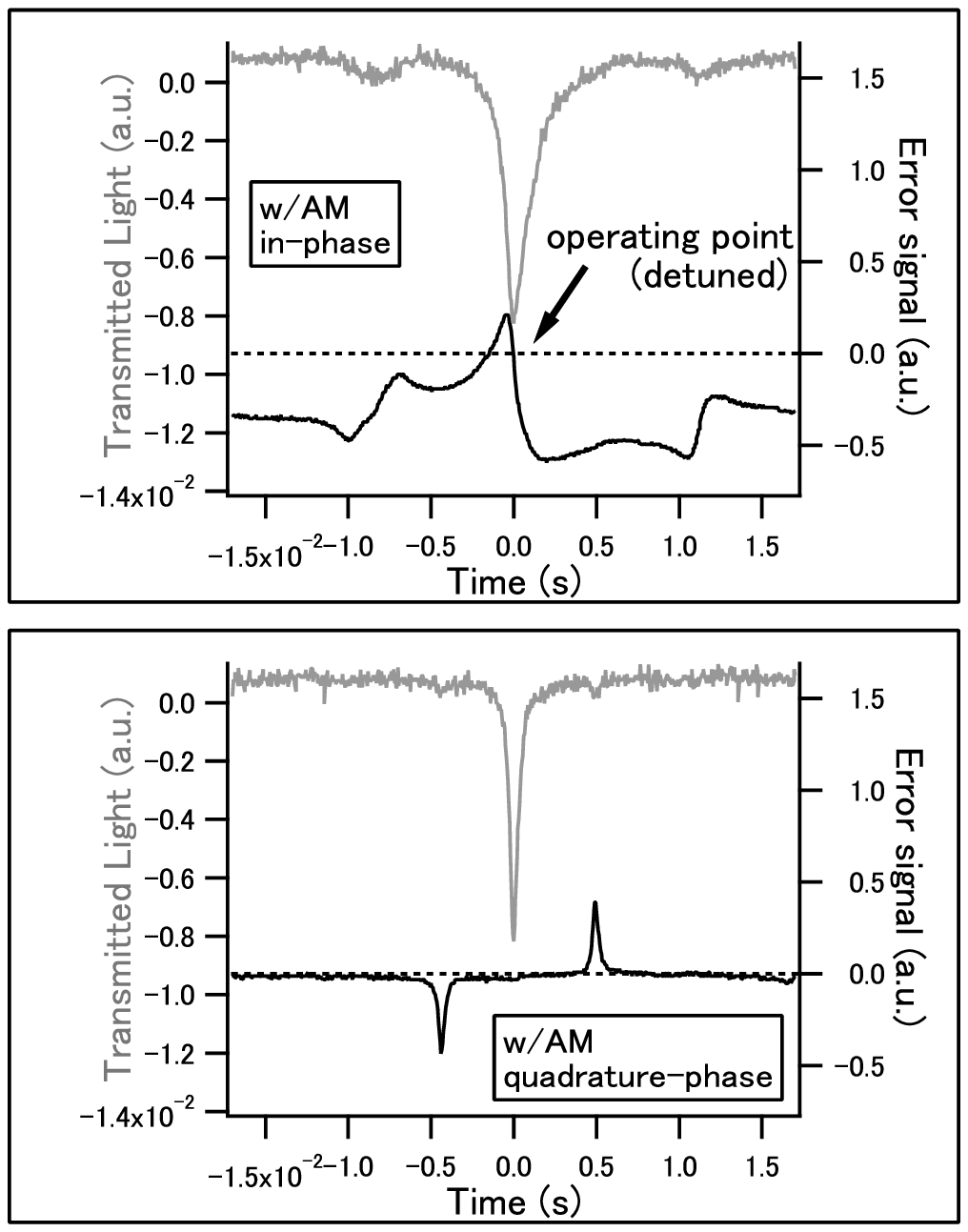}
  \caption{Error signals in in-phase (bottom) and in quadrature-phase (top) with (right panels) and without (left panels) the AM sidebands. While there is no difference in the quadrature-phase signals, the in-phase signal has an offset when AM modulation is applied. The transmitted light increases without the AM sidebands as the overall modulation depth has been reduced by changing the polarization of the light injected to the EOM.}
 \label{fig:data}
\end{figure}
\end{center}

Demodulating the output signal of the photo-detector with a proper demodulation phase, one obtains the Pound-Drever-Hall error signal~\cite{PDH} that is a zero-crossing around the cavity resonance and gives a linear response to the motion of the steering mirror. Figure~\ref{fig:data} shows the error signal and the transmitted light of the cavity. Here, the steering mirror' position is swept via a piezo actuator.
 In the left panels, the half-wave plate is tuned in a way that the modulation is purely a phase modulation. In the right panels, the half-wave plate is tuned to add a fraction of amplitude modulation. In this way, the relative modulation phase of the AM to PM is almost zero, which means the AM does not create an imbalance between the upper and lower sidebands but only introduces a rotation in the phasor diagram. In fact, there is no need of the imbalance in our experiment as we do not have a Michelson interferometer. In the left panels, the error signal neither has an offset in the in-phase nor the quadrature-phase.
 In the right panels with the AM, the in-phase error signal receives an offset and the zero-crossing point is shifted slightly to the left because of the use of the AM sidebands. Feeding back the error signal to the mirror, one can lock the cavity with some detuning. Note that the error signal has no offset in the quadrature-phase even with the AM. In the full RSE interferometer, we need two independent modulators to tune both the AM modulation depth and the relative phase to the PM.

\begin{center}
\begin{figure}[htbp]
  \includegraphics[width=7cm]{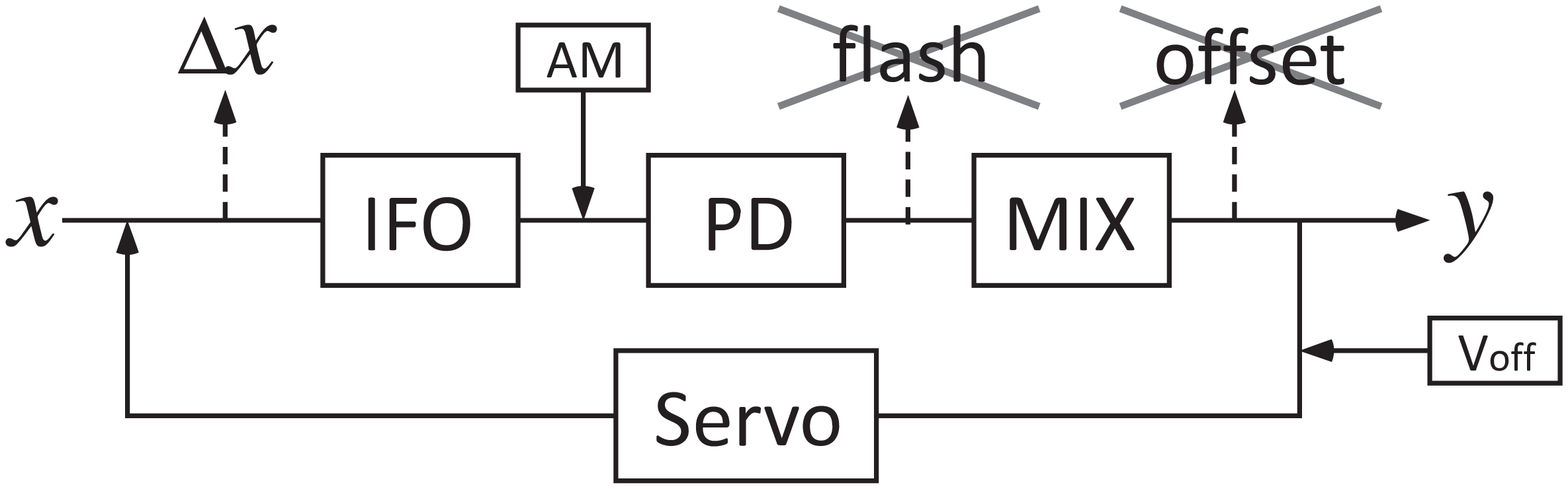}
\hspace{0.5cm}
  \includegraphics[width=7cm]{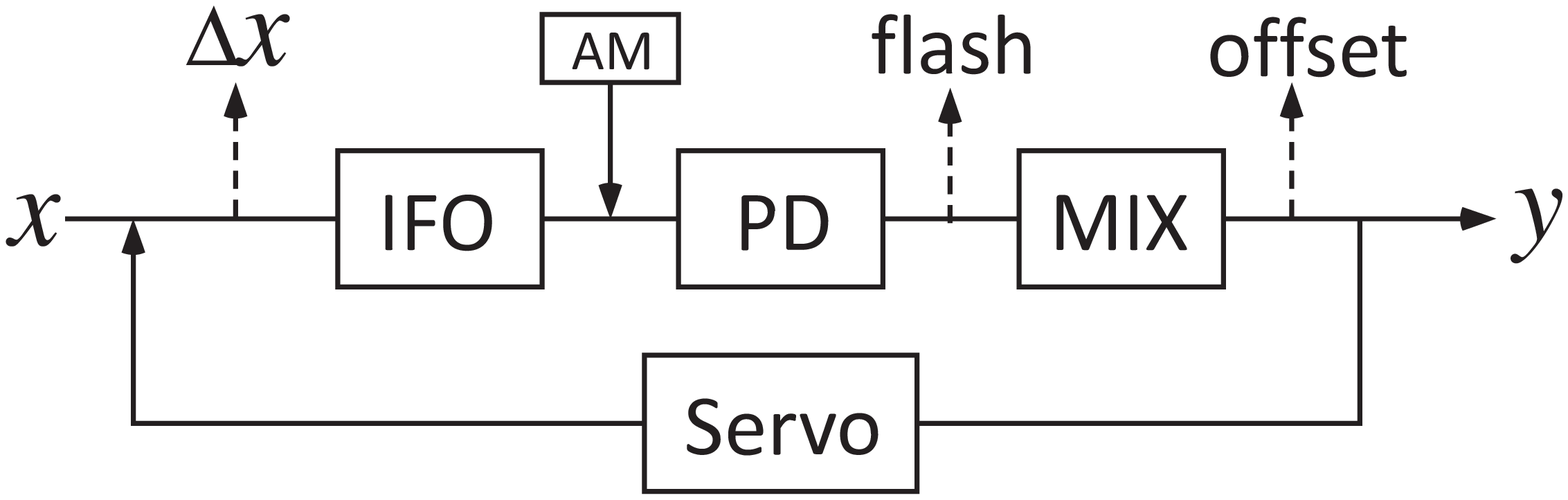}
  \caption{Block diagrams of the interferometer control system for KAGRA (left) and our simple experiment (right). In both cases, the mirror motion $x$ is probed by an interferometer (IFO), converted to an electric voltage at the photo-detector (PD), and demodulated by a modulation frequency at the mixer (MIX) to create an output signal $y$, which is fed back to the mirror via a control servo.}
 \label{fig:servo}
\end{figure}
\end{center}

Figure~\ref{fig:servo} shows block diagrams of the servo systems for the KAGRA interferometer with the full RSE configuration (left panel) and our simple experiment with a single cavity (right panel). In the case of KAGRA, an offset voltage is added to the error signal before the feedback. Defining $H$ as the interferometer optical gain, $G$ as the servo gain, $V_\mathrm{off}$ as the offset voltage, and $V_\mathrm{AM}$ as the negative offset voltage added by the AM sidebands, one obtains the shift of the signal-recycling mirror position $\Delta x$ and the output signal $y$ to be 
\begin{eqnarray}
\Delta x&=&\frac{1}{1+GH}\left[x-G\left(V_\mathrm{off}+V_\mathrm{AM}\right)\right]\ ,\label{eq:deltax}\\
y&=&\frac{1}{1+GH}\left[Hx-GHV_\mathrm{off}+V_\mathrm{AM}\right]\ .
\end{eqnarray}
In order to cancel the offset at the output $y$, the AM voltage $V_\mathrm{AM}$ shall be tuned to $GHV_\mathrm{off}$. As a result, Eq.~(\ref{eq:deltax}) reads
\begin{eqnarray}
\Delta x&\rightarrow&\frac{x}{1+GH}-GV_\mathrm{off}\ ,
\end{eqnarray}
which indicates that the mirror be locked to the detuned position with $GH$ sufficiently higher than unity at DC.

In the case of our experiment, the offset voltage is not added for simplicity. The steering mirror position $\Delta x$ and the output signal $y$ are derived to be
\begin{eqnarray}
\Delta x&=&\frac{1}{1+GH}\left[x-GV_\mathrm{AM}\right]\ ,\label{eq:deltax2}\\
y&=&\frac{1}{1+GH}\left[Hx+V_\mathrm{AM}\right]\ .
\end{eqnarray}
The output signal is offset by the additional AM and the operation point is shifted as shown in Fig.~\ref{fig:data}. 

\section{Simulation} \label{sec:6}

Let us now quantitatively evaluate the effect of the additional AM sidebands on the noise reduction by simulations.
We use an interferometer simulation code {\it Optickle}~\cite{optickle} with the parameters for KAGRA (Table~\ref{tb:prKAGRA}). The optical parameters are chosen to optimize the observation range for neutron star binaries as well as a broad bandwidth to cover high-frequency gravitational-wave signals~\cite{KAGRA}. The input carrier power is limited by the cooling capability of the cryogenic system. The input laser power in the fundamental Gaussian mode including the sidebands fields is as high as 100~W, so we have some extra laser power that can be used to generate the sidebands fields. The modulation depth $m$ is determined not to degrade the sensitivity by the control loop noise~\cite{LSC}.
\begin{table}[htbp]
\begin{center}
\begin{tabular}{|c|c|}\hline
Item&Value\\ \hline \hline
arm cavity finesse&1550\\
power/signal-recycling cavity length & 66.591\,m \\
signal-recycling cavity detune phase & 3.5\,deg\\
input carrier power & 82\,W\\
1st PM frequency & 16.880962\,MHz\\
2nd PM frequency & 45.015898\,MHz\\
PM modulation depth & 0.1\,rad\\
optical loss of test mass & 45\,ppm\\
beamsplitter reflectivity/transmittance& 50.5\,\% / 49.5\,\% \\
transmittance of end test mass&10\,ppm\\
power/signal-recycling mirror reflectivity&90\,\%/ 85\,\%\\ 
resolution of photo detector & $1\mathrm{nV}/\sqrt{\mathrm{Hz}}$ for 100\,mV \\
oscillator single-sideband phase noise & -120\,$\mathrm{dBc}/\sqrt{\mathrm{Hz}}$ \\\hline\hline
Degree of freedom & Extraction port \\ \hline \hline
CARM & 45\,MHz, in-phase, REFL \\
DARM & DC, AS \\
PRCL & 45\,MHz, in-phase, POP\\
MICH & 16\,MHz, quadrature-phase, POP \\
SRCL & 16\,MHz, in-phase, POP \\ \hline
\end{tabular}
\end{center}
\caption{Interferometer parameters of KAGRA and proposed signal extraction ports for the length control.}
\label{tb:prKAGRA}
\end{table}

\begin{figure}[htbp]
\begin{tabular}{cc}
 \begin{minipage}{0.5\hsize}
 \begin{center}
  \includegraphics[width=8cm]{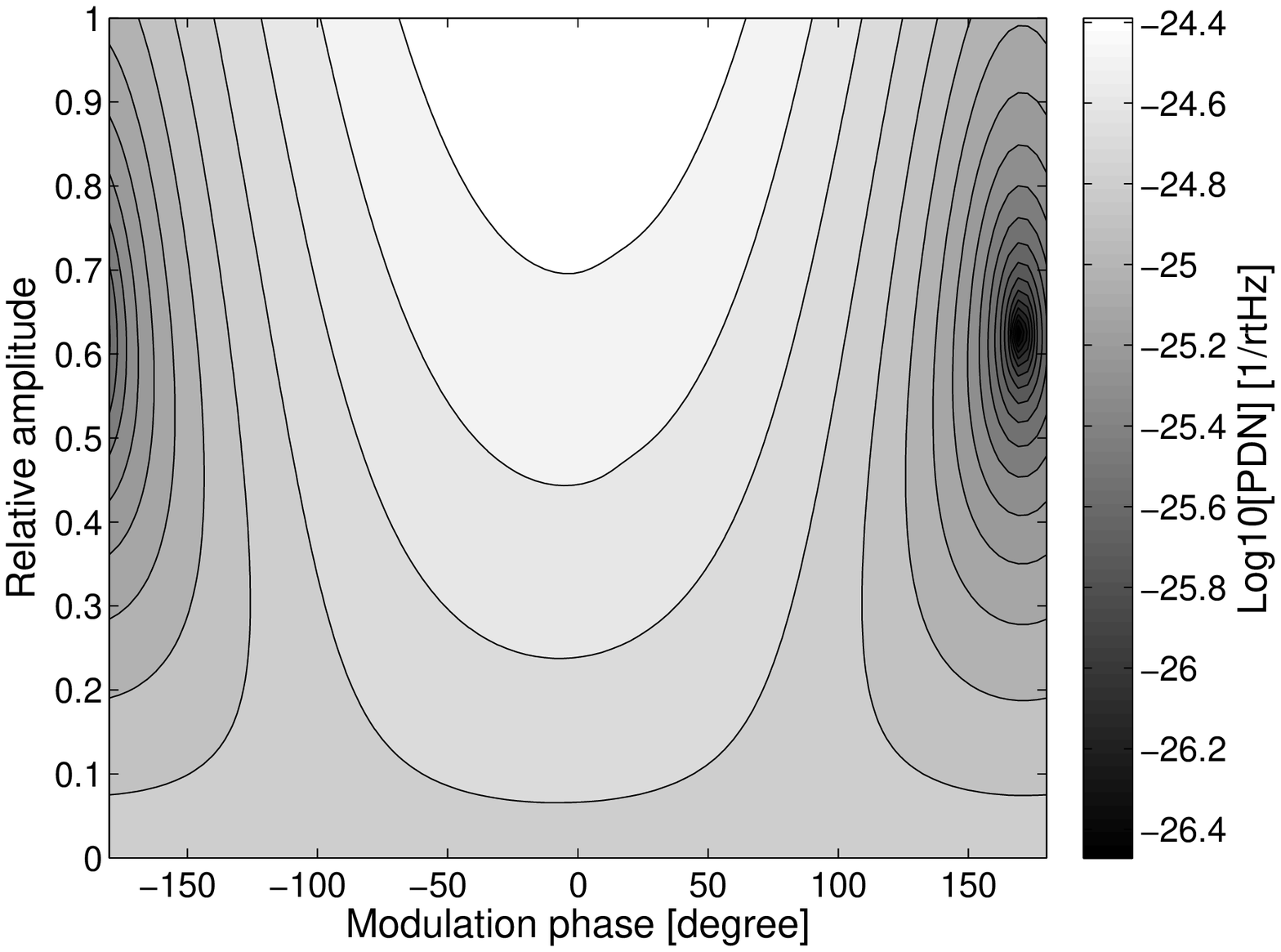}
 \end{center}
 \end{minipage}
 \begin{minipage}{0.5\hsize}
 \begin{center}
  \includegraphics[width=8cm]{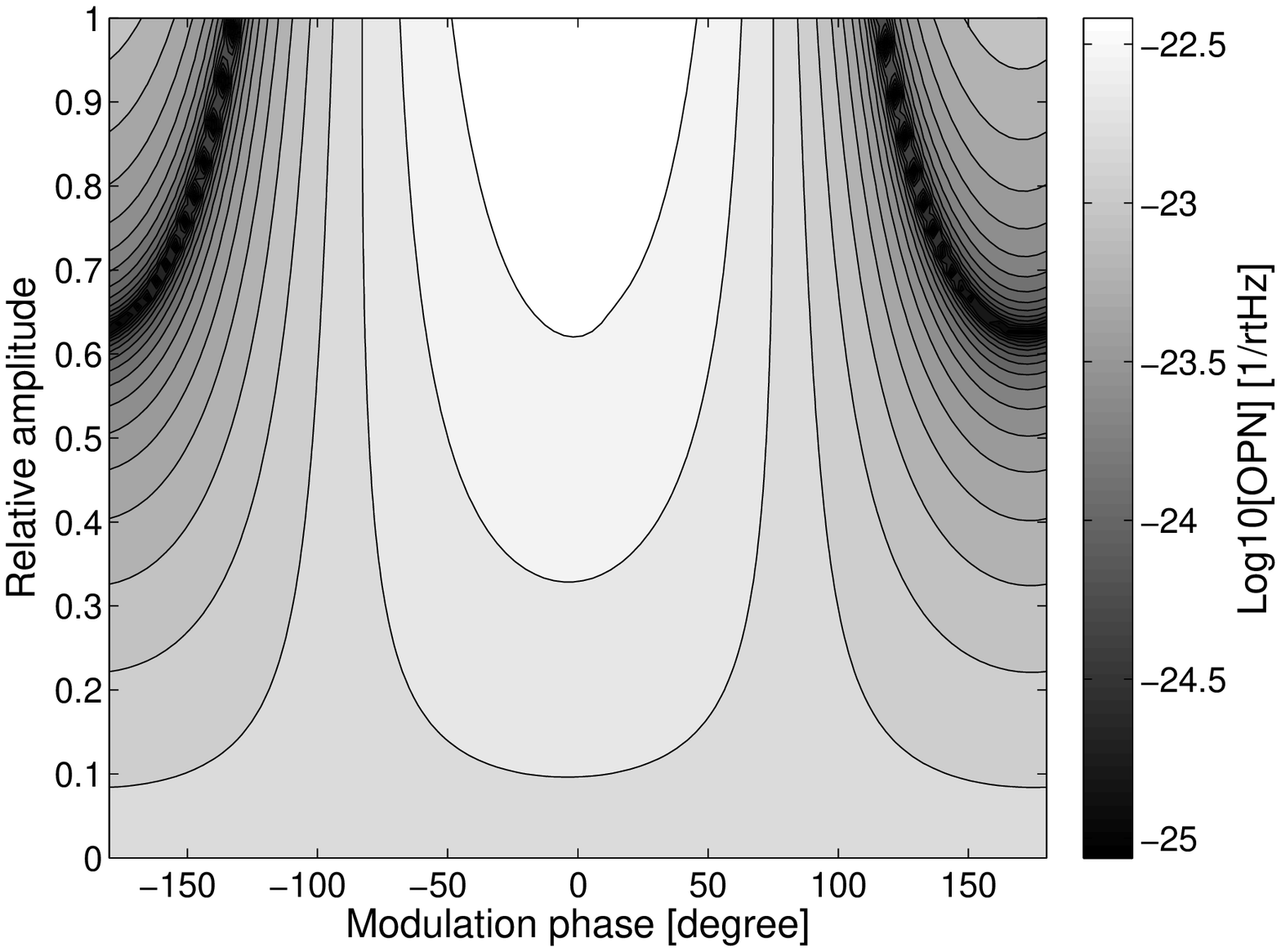}
 \end{center}
  \end{minipage}
 \end{tabular}
 \caption{Contour plot of PDN (left) and OPN (right) with different relative amplitudes and relative modulation phases of the AM sidebands. The noise levels are logarithmically shown in $1/\sqrt{\mathrm{Hz}}$.}
 \label{contourPlot}
\end{figure}

\begin{figure}[htbp]
 \begin{center}
  \includegraphics[width=16cm]{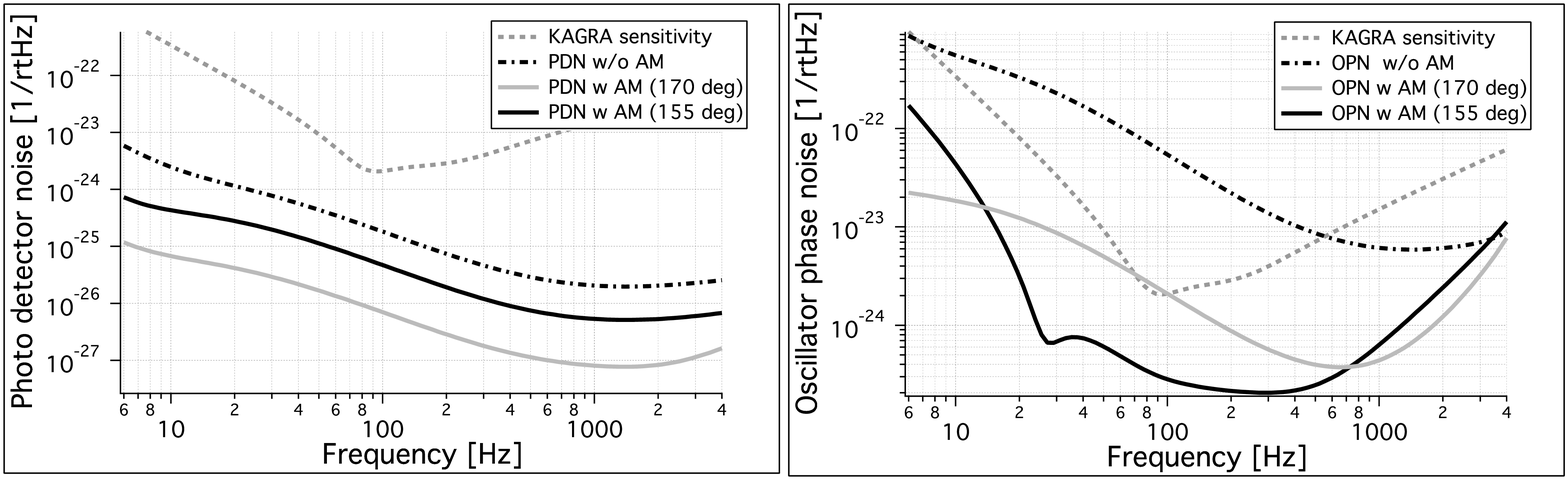}
 \end{center}
 \caption{{\it Left}: PDN spectrum. The gray dashed line is the target sensitivity of KAGRA. The black dash-dotted line is PDN without the AM sidebands. The black and gray solid lines are PDN with the AM sidebands where the modulation amplitude and phase are chosen to minimize OPN and PDN, respectively. {\it Right}: OPN spectrum. The indices of the lines are the same as those in the left panel.}
 \label{PDNandOPN}
\end{figure}

\begin{figure}[htbp]
 \begin{center}
  \includegraphics[width=7cm]{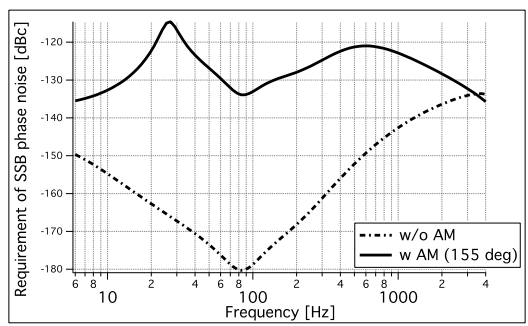}
 \end{center}
 \caption{The requirement of the phase fluctuation of the oscillator. The black dash-dotted line shows the requirement in the detuned RSE. The solid line is the requirement using our method with the AM relative modulation phase of 155\,deg.}
 \label{reqosci}
\end{figure}

The purpose of the simulation is as follows: (i) verification of the reduction of PDN and OPN with the additional AM sidebands, (ii) evaluation of the required modulation depth and the modulation phase of the AM sidebands, and (iii) investigation of other noise possibly introduced by adding the AM sidebands. 
The PDN level is estimated simply by calculating the amount of the offset light at 16\,MHz (see Eq.~(ref:PDN)) with Optickle. 
The OPN level is estimated in the following way. First we calculate transfer functions from the phase modulation of the RF oscillator to the DARM channel and from the DARM modulation to the DARM channel with Optickle. Second we take the ratio of the two transfer functions. At last we multiply a typical value of oscillator fluctuation to the ratio.
Both the amount of the offset and the transfer function from the RF oscillator to the DARM channel depend on the amplitude and the modulation phase of the 16\,MHz AM sidebands relative to the PM sidebands. We shall sweep the relative amplitude and phase to search the optimal condition to reduce the noise levels. Figure~\ref{contourPlot} shows the noise levels in contour plots. Here the measurement frequency is fixed to 100\,Hz. One finds a significant reduction of the noise levels at around a certain combination of the relative amplitude and phase. It turns out that the optimal combinations to minimize PDN and to minimize OPN are slightly different.
The optimal pair to minimize PDN is 65\,\% and 170\,deg while the optimal pair to minimize OPN is 65\,\% and 155\,deg. The difference results from the fact that OPN appears in the length sensing signals of other degrees of freedom and couple to the gravitational-wave channel via a number of paths while PDN is local sensing noise. With the optimum values, the PDN and OPN levels decrease remarkably as it is shown in Fig.~\ref{PDNandOPN}. Here, we assume that the oscillator single-sideband phase noise level is -120\,dBc/$\sqrt{\mathrm{Hz}}$, the photo-detector output voltage noise is $1\mathrm{nV}/\sqrt{\mathrm{Hz}}$, and the maximum photo-detector output is 100\,mV. With these values, OPN is more serious than PDN. Applying the relative amplitude and phase that are optimal for OPN (65\,\% and 155\,deg), we obtain the results with both PDN and OPN below the KAGRA sensitivity, thus it is better to focus on OPN. Instead of calculating the noise spectrum with a fixed oscillator noise level, we can calculate the requirement to the oscillator that is to be installed in KAGRA. The result is shown in Fig.~\ref{reqosci}.

Reducing the relative amplitude of the AM sidebands is beneficial to save the total laser power, but the OPN level increases rapidly. We shall calculate how much the relative amplitude can be reduced to satisfy the requirement of KAGRA.
 The optimum combination is, as we have seen, 65\,\% and 155\,deg, and OPN decreases by a factor of 192 at 100\,Hz compared with the noise level without the additional AM sidebands. With the relative amplitude of 60\,\%, for example, the optimal relative phase is then 172\,deg, and OPN decreases only by a factor of 15. In order to suppress OPN 10 times below the sensitivity curve of KAGRA, the lowest value of the relative amplitude is 63\,\%.

Generation of an AM sidebands accompany the power loss of the carrier light unless an additional interferometer is used~\cite{Ohmae}, so that it would be impractical if the required AM sidebands amplitude is too high. A typical amplitude modulator consists of two electro-optic crystals in series with the crystal axes rotated by 90\,deg around the optical axes to compensate for any change in birefringence owing to temperature change. The output of the amplitude modulator is~\cite{Ohmae}
\begin{eqnarray}
E(t)&\simeq&E_0J_0(m)\cos{\phi_0}\cdot e^{\mathrm{i}\Omega t}+\mathrm{i}E_0J_1(m)\sin{\phi_0}\left[e^{\mathrm{i}(\Omega+\omega) t}-e^{\mathrm{i}(\Omega+\omega) t}\right],
\end{eqnarray}
where $E_0$ is the amplitude of the input field and $\phi_0$ is the bias phase of the operating point of the modulator. Since the carrier power of KAGRA to be injected to the interferometer can be at most 100\,W and the required carrier power is 82\,W, the modulation depths and the bias phase should satisfy (i) $J_0(m)^4[J_0(m')\cos{\phi_0}]^2\geq0.82$. Here $m'$ is the modulation depth on the electro-optic crystal in the amplitude modulator. For the amplitude of the AM sidebands to be 65\,\% of the PM sidebands, the modulation depths and the bias phase should also satisfy (ii) $J_1(m)\cos{\phi_0}\times 0.65=J_2(m')\sin{\phi_0}$. The conditions (i) and (ii) can be satisfied with $m<0.25$. With $m=0.1$, for example, the total power loss by the two EOMs and one amplitude modulator is as low as 5\,\%, in which case $m'=0.16$ and $\phi_0=0.013$~rad.

\begin{figure}[t]
 \begin{center}
  \includegraphics[width=14cm]{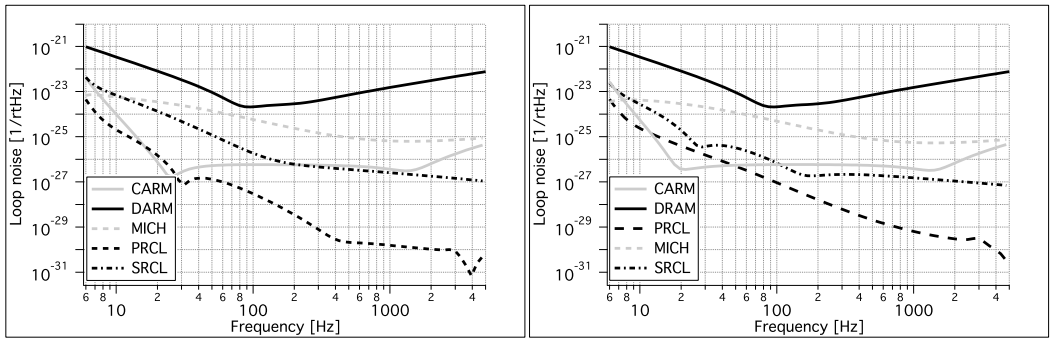}
 \end{center}
 \caption{{\it Left}: Control loop noise of KAGRA without the additional AM sidebands. {\it Right}: Control loop noise of KAGRA with additional AM sidebands.}
 \label{fig:loopnoise}
\end{figure}

At last, we shall consider if sensing noise on other control signals could increase with the additional AM sidebands. We have calculated control loop noise using Optickle. It includes shot noise of all the 5 degrees of freedom that can couple to the gravitational-wave channel via the control servo loop~\cite{LSC}. The control bandwidth is set to 20~Hz. Figure~\ref{fig:loopnoise} shows the result. There is no remarkable change in MICH loop noise. The CARM and PRCL has increased a little but they do not affect the sensitivity of KAGRA. 

\section{Summary} \label{sec:7}

We propose a method to use an additional amplitude modulation sidebands to remove a non-trivial excess noise in a detuned RSE interferometer. The excess noise sources are the reduced photo-detector resolution and the fluctuation of the oscillator for the control sidebands. Our simulation results show a significant reduction of excess noise. The AM modulation depth and the relative modulation phase to the PM sidebands have to be tuned to minimize the excess noise level. We have found that the optimum relative phases are slightly different between the minimization of photo-detector noise and the minimization of oscillator phase noise, but the overall noise level is well below the KAGRA requirement by taking the optimum for the latter. Generation of the AM sidebands accompanies a loss of the carrier power, but the total loss at all the three modulators to minimize excess noise with the detuning turns out to be as low as 5\,\% for KAGRA.
We have also demonstrated the method in a simple experiment with adding an amplitude modulation to shift the operating point of a cavity. We concluded that the detuning technique can be used in the second-generation gravitational-wave detectors.

\appendix
\section{Solution for the unbalanced tilted sidebands}\label{sec:cal}
It is important to show that our application is valid even if the PM sidebands are both unbalanced in the amplitude and tilted in the phasor diagram. The electric field reflected from the detuned interferometer with the additional AM sidebands are described as
\begin{eqnarray}
E &= E_0^{\mathrm{D}} \mathrm{e}^{\mathrm{i}\Omega t} + \left(\mathrm{i}E_{+1}^{\mathrm{D}}\mathrm{e}^{\mathrm{i}(\Omega+\omega) t} +\mathrm{i} E_{-1}^{\mathrm{D}}\mathrm{e}^{\mathrm{i}(\Omega-\omega) t}\right. \nonumber \\
&\hspace{30mm}\left.+ E_{+1}^{\mathrm{AM}}\mathrm{e}^{\mathrm{i}((\Omega+\omega) t + \beta)} +E_{-1}^{\mathrm{AM}}\mathrm{e}^{\mathrm{i}((\Omega-\omega) t - \beta)}\right)\mathrm{e}^{\mathrm{i} \alpha}. \label{wAM}
\end{eqnarray}
The conditions to make the summed sidebands field a pure PM sidebands to get rid of PDN are: (i) the real parts of the summed upper and lower sidebands amplitudes are of equal magnitude but opposite signs, and (ii) the imaginary parts of the summed upper and lower sidebands amplitudes are equal. The conditions (i) and (ii) read
\begin{eqnarray}
(1-a)\sin{\alpha}E_{\pm1}^\mathrm{D} &= \left[\cos{(\alpha-\beta)}-a\cos{(\alpha+\beta)}\right]E_{\pm1}^\mathrm{AM}\ , \label{eq:conPDN1}\\
(1-a)\cos{\alpha}E_{\pm1}^\mathrm{D} &= -\left[\sin{(\alpha-\beta)}-a\sin{(\alpha+\beta)}\right]E_{\pm1}^\mathrm{AM}\ , \label{eq:conPDN2}
\end{eqnarray}  
respectively.

The conditions to cancel out OPN can be obtained in the following way. We shall add a small phase fluctuation $\Delta\phi$ to Eq.~(\ref{wAM}) and take its first order:
\begin{eqnarray}
E &\simeq E_0^{\mathrm{D}}\mathrm{e}^{\mathrm{i}\Omega t} + \left[\mathrm{i}E_{+1}^{\mathrm{D}}\mathrm{e}^{\mathrm{i}(\Omega+\omega) t}(1 + \mathrm{i}\Delta\phi) + \mathrm{i}E_{-1}^{\mathrm{D}}\mathrm{e}^{\mathrm{i}(\Omega-\omega) t}(1 - \mathrm{i}\Delta\phi)\right.\nonumber \\
&\hspace{5mm} + \left.E_{+1}^{\mathrm{AM}}\mathrm{e}^{\mathrm{i}((\Omega+\omega) t + \beta)}(1 + \mathrm{i}\Delta\phi) + E_{-1}^{\mathrm{AM}}\mathrm{e}^{\mathrm{i}((\Omega-\omega) t - \beta)}(1 - \mathrm{i}\Delta\phi)\right]\mathrm{e}^\mathrm{i\alpha}\ .\label{eq:totalE}
\end{eqnarray}
The conditions are: (i)' the real parts of the summed upper and lower OPN sidebands are of equal magnitude but opposite signs, and (ii)' the imaginary parts of the summed upper and lower OPN sidebands are equal. The conditions (i)' and (ii)' read
\begin{eqnarray}
(1-a)\cos{\alpha}E_{\pm1}^\mathrm{D} &= -\left[\sin{(\alpha-\beta)}-a\sin{(\alpha+\beta)}\right]E_{\pm1}^\mathrm{AM}\ , \label{eq:conOPN1}\\
(1-a)\sin{\alpha}E_{\pm1}^\mathrm{D} &= \left[\cos{(\alpha-\beta)}-a\cos{(\alpha+\beta)}\right]E_{\pm1}^\mathrm{AM}\ , \label{eq:conOPN2}
\end{eqnarray}  
respectively.

Since the conditional-equation pair to minimize OPN (Eqs.~(\ref{eq:conPDN1}) and (\ref{eq:conPDN2})) is equivalent to the pair to minimize PDN (Eqs.~(\ref{eq:conOPN1}) and (\ref{eq:conOPN2})), both OPN and PDN can be removed simultaneously.
\\

\end{document}